\documentclass[preprint,showpacs]{revtex4}

\usepackage[latin1]{inputenc}
\usepackage{graphicx}%
\usepackage{dcolumn}
\usepackage{amsmath}

\makeatletter
\def\btt#1{\texttt{\@backslashchar#1}}%
\DeclareRobustCommand\bblash{\btt{\@backslashchar}}%
\makeatother

\begin{document}


 \title{$\lq\lq$Cumulated Vehicle Acceleration": An Attribute
 of   GPS Probe Vehicle Traces for On-Line Assessment of Vehicle Fuel Consumption in Traffic and Transportation Networks}

\mark{$\lq\lq$Cumulated  Vehicle Acceleration"}

\author{   
Boris S. Kerner$^1$}

 \affiliation{$^1$
Physik von Transport und Verkehr, Universit{\"a}t Duisburg-Essen,
47048 Duisburg, Germany}



\pacs{89.40.-a, 47.54.-r, 64.60.Cn, 05.65.+b}

\begin{abstract} 
 To perform a reliable on-line assessment of   fuel consumption in vehicles,
 we introduce $\lq\lq$cumulated vehicle acceleration" as an attribute
 of GPS probe vehicle traces.  
The objective of the calculation of the attribute $\lq\lq$cumulated vehicle acceleration" in the GPS probe vehicle data
  is  to perform a reliable on-line dynamic traffic assignment for
 the reduction of vehicle consumption in traffic and transportation networks.
\end{abstract}

\maketitle

\section{Introduction}

A reliable assessment of   fuel consumption in vehicles is an important task of transportation engineering (see, for 
example~\cite{Akcelik,Barth,Minett,Boriboonsomsin,Bandeira,Ahn,Rakha,Frey,Barth2}
and references there).
It is well-known that at a time-independent speed the speed dependence of fuel consumption in a vehicle has a minimum
at speeds about 50--70 km/h. Therefore, an average speed is an usual
attribute for a reliable assessment of   fuel consumption in vehicles from measurements of GPS probe vehicle traces (see, for 
example~\cite{Akcelik,Barth,Minett,Boriboonsomsin,Bandeira,Ahn,Rakha,Frey,Barth2}
and references there).

However, it is also well-known that fuel consumption of a vehicle depends crucially on vehicle acceleration: Usually, at a given   speed, the larger the 
vehicle acceleration, the larger the fuel consumption of the vehicle. 
There are many methods in which through the use of measurements of engine characteristics together
with simulations of driver behavior, fuel consumption as a function of speed and acceleration (deceleration) has been calculated (see, for 
example~\cite{Akcelik,Barth,Minett,Boriboonsomsin,Bandeira,Ahn,Rakha,Frey,Barth2}
and references there).

Recently, based on a huge number of measurements of fuel consumption in floating car data (FCD),
Koller et al.~\cite{Koller} have found a $\lq\lq$pure" empirical microscopic matrix: The matrix has been derived with the use of measured data only, i.e.,
without any
models and simulations. This empirical matrix presents microscopic empirical fuel consumption as a function of microscopic (single-vehicle) 
speed and acceleration (deceleration).
This matrix can be used for many ITS (intelligent transportation systems) applications~\cite{Hemmerle2,Hemmerle3}.

However, due to an error in the determination of the   vehicle location   in GPS probe vehicle traces, it is almost impossible to determine
real vehicle acceleration as a time-function with measurements of the GPS   traces.
Therefore, additionally to the attribute $\lq\lq$average vehicle speed", for an assessment of  vehicle fuel consumption on a link of a traffic network
researches have suggested several attributes, which should be found with the use of the GPS   traces, like the
number of vehicle stops on a link of a traffic network, kinetic energy of vehicles, etc.

In this paper,   we introduce  an attribute of GPS probe vehicle data that we call $\lq\lq$cumulated vehicle acceleration".  
 $\lq\lq$Cumulated vehicle acceleration"
together with well-known attribute of GPS probe vehicle data $\lq\lq$average vehicle speed" should allow us to   perform a reliable assessment of  vehicle fuel consumption on a link of a traffic network.

\section{Definition of $\lq\lq$Cumulated Vehicle Acceleration"}

\begin{figure}
\begin{center}
\includegraphics*[width=10 cm]{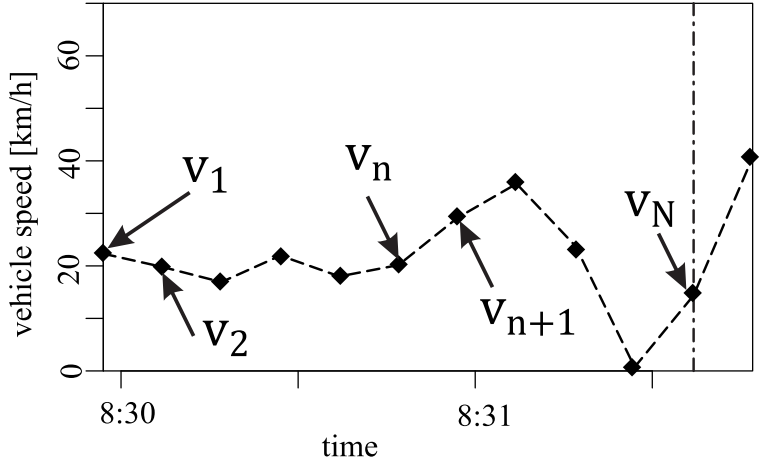}
\end{center}
\caption[]{Typical microscopic (single-vehicle) speeds (black
squares) along a vehicle trajectory measured through anonymized GPS probe vehicle data of TomTom company. 
Dash-dotted line shows   time
instance of vehicle passing traffic signal;
the signal location is considered   the end of the city link.
Data
have been measured on
  a section of V{\"o}lklinger Stra{\ss}e with speed limit 60 km/h (see schema of this section in Fig.~1 (a)
   of Ref.~\cite{Kerner_City}). 
  }
\label{Speed}
\end{figure}

In Fig.~\ref{Speed}, we present a typical dependence of microscopic vehicle speeds determined through anonymized GPS   traces of TomTom company.
Features of these measured data have been discussed in detail in~\cite{Kerner_City}.

 We define a $\lq\lq$cumulated vehicle acceleration" $A_{j}$ for link $j$  of a traffic network  as follows~\cite{Kerner}:
 \begin{equation}
 A_{j}= \frac{1}{L_{j}}\sum^{N-1}_{n=1}(v_{n+1}-v_{n})\theta_{n},
 \label{cumulated_F}
 \end{equation}
 where $L_{j}$ is the length of link $j$ of the traffic network, $v_{n}$ is   a microscopic speed measured 
 along vehicle trajectory on link $j$ of the traffic network at time instant $n$  (here $n= 1, 2, ..., N$, i.e., $v_{1}$ is the first value of a microscopic speed measured 
 along vehicle trajectory on link $j$ of the traffic network, $v_{N}$ is the last value of a microscopic speed measured 
 along vehicle trajectory on link $j$ of the traffic network, see Fig.~\ref{Speed}), 
\begin{eqnarray}
\theta_{n}=\left\{
\begin{array}{ll}
1 & \textrm{if $v_{n+1}-v_{n} \geq \Delta v$} \\
0 &  \textrm{if $v_{n+1}-v_{n} < \Delta v$},
\end{array} \right.
\label{theta}
\end{eqnarray}
$\Delta v$  is a constant model parameter ($\Delta v>0$) that is used to decrease the effect of an error
in the vehicle speed $v_{n}$ calculated from GPS   data; 
for anonymized TomTom GPS data with time interval $\Delta t=t_{n+1}-t_{n}=$ 5 s (see explanations in~\cite{Kerner_City}) between measurements of the speeds
 one can use, for example,    $\Delta v=$ 0.5 [km/h].

It should be noted that the definition of $\lq\lq$cumulated vehicle acceleration" $A_{j}$ for link $j$ of a traffic network
can be applied for any link  of a traffic network independent on the length of the link. Additionally, 
the definition of $\lq\lq$cumulated vehicle acceleration" $A_{j}$  (\ref{cumulated_F}) for link $j$ of the traffic network
can be applied for any value of 
  time interval $\Delta t_{n}=t_{n+1}-t_{n}$ (where $n= 1, 2, ..., N-1$) between   GPS measurements.
The latter is associated with definition  (\ref{cumulated_F}), in which
the sum of  speed differences
between any two measurements of GPS-points is calculated.   

The basic assumption made in  definition  (\ref{cumulated_F}) of $\lq\lq$cumulated vehicle acceleration" $A_{j}$ for link $j$ of a traffic network
is as follows. We  assume that vehicle acceleration $a_{n}$  is time-independent during
time interval $\Delta t_{n}$, i.e., it is equal to
 \begin{equation}
 a_{n}= \frac{v_{n+1}-v_{n}}{\Delta t_{n}},
 \label{a_F}
 \end{equation}
 where  $n= 1, 2, ..., N-1$.
 
 \section{Discussion}
 
 Recently, Hemmerle et al.~\cite{Hemmerle} and Hermanns et al.~\cite{Hermanns} have calculated a macroscopic matrix for
  fuel consumption on links of a traffic network.  In this matrix, average fuel consumption on a link of traffic network depends
  on two variables: (i) $\lq\lq$average vehicle speed"
 and (ii) $\lq\lq$cumulated vehicle acceleration"  as  defined   in this paper. 
 
 To calculate the macroscopic
 fuel consumption matrix, firstly, a classification of simulated traffic patterns
 in oversaturated city traffic has been made by Hermanns et al.~\cite{Hermanns}.
 For each of the traffic patterns found in microscopic traffic simulations with Kerner-Klenov stochastic three-phase traffic flow model
   (about 500000 patterns have been calculated)
 the average speed and the cumulated vehicle acceleration have been found~\cite{Hemmerle,Hermanns}.
 Finally, with the use of the empirical microscopic fuel consumption matrix of Koller et al.~\cite{Koller},
 Hemmerle et al.~\cite{Hemmerle} have derived the macroscopic fuel consumption matrix. As mentioned, in the matrix
 average fuel consumption on a link of traffic network depends
  on the average vehicle speed 
 and the cumulated vehicle acceleration.
 
 As shown in paper by Hemmerle et al.~\cite{Hemmerle},  with the use of this matrix, it is possible to perform 
 {\it on-line} dynamic traffic assignment based on actual   GPS probe vehicle traces.
 In accordance with the methodology developed in German project UR:BAN~\cite{URBAN,Rehborn},
 the objective of such on-line dynamic traffic assignment is the reduction of vehicle consumption in city networks.
 
 \section{Conclusions}
 
1. $\lq\lq$Cumulated vehicle acceleration" as defined in this paper is an useful variable for on-line
assessment of vehicle fuel consumption based on measurements of GPS probe vehicle data.

2. Through the calculation of $\lq\lq$cumulated vehicle acceleration" in measured GPS probe vehicle data,
  a reliable on-line dynamic traffic assignment is possible to perform. The objective of such on-line dynamic traffic assignment
is the reduction of vehicle consumption in traffic and transportation networks.
 
 {\bf Acknowledgment:}
We thank our partners for their support in the project $\lq\lq$UR:BAN -
    Urban Space: User oriented assistance systems and network management",
    funded by the German Federal Ministry of Economic Affairs and Energy.


\begin{thebibliography}{8.}
\addcontentsline{toc}{section}{References}


 
 
 \bibitem{Akcelik}
 Akcelik, R., C. Bailey, C. Bowyer, and D.C. Biggs. A Hierarchy of Vehicle Fuel Consumption Models.   Traffic Engineering and Control, Vol. 24, No. 10, 1983, pp. 491--495.
 
 \bibitem{Barth}
 	
Barth, M., F. An, J. Norbeck, and M. Ross. Modal emissions modeling: A
      physical approach. In Transportation Research Record: Journal of the
      Transportation Research Board, {\bf 1520},     81--86 (1996).	
      
       
\bibitem{Minett}

Minett, C.F., A.M. Salomons, W. Daamen, B. von Arem, and S. Kuijpers. 	Eco-routing: comparing the fuel consumption of different routes between an 	origin and destination using field test speed profiles and synthetic speed profiles. 	In Proceedings 2011 IEEE Forum on Integrated and Sustainable Transportation 	Systems, Vienna, Austria, 2011. 
   
   \bibitem{Boriboonsomsin}
  Boriboonsomsin, K., M. J. Barth, W. Zhu, and A. Vu. Eco-Routing Navigation System Based on Multisource Historical and Real-Time Traffic Information. 
  Intelligent Transportation Systems, IEEE Transactions, {\bf 13}, No. 4,   1694--1704 (2012).

\bibitem{Bandeira}    
 Bandeira, J. M., T. G. Almeida, A. J. Khattak, N. M. Rouphail, and M. C. Coelho. Generating Emissions Information for Route Selection: Experimental 
 Monitoring and Routes Characterization. Journal of Intelligent Transportation Systems, {\bf 17}, No. 1,     3--17 (2013).
\bibitem{Ahn}      
 Ahn, K., and H. Rakha. Field evaluation of energy and environmental impacts of driver route choice decisions. 
 2007 IEEE Intelligent Transportation Systems Conference, Vol. 1 and 2,   pp. 216--221 (2007).
\bibitem{Rakha}   
 Rakha, H., K. Ahn, and A. Trani. Development of VT-Micro model for estimating hot stabilized light 
 duty vehicle and truck emissions. Transportation Research Part D: Transport and Environment, {\bf 9}, No. 1,  49--74 (2004).
\bibitem{Frey}   
  Frey, H. C., K. S. Zhang, and N. M. Rouphail. Fuel use and emissions comparisons for alternative routes, 
  time of day, road grade, and vehicles based on in-use measurements. Environmental Science and Technology, {\bf 42}, No. 7,   2483--2489 (2008).
\bibitem{Barth2}  
   Barth, M., and K. Boriboonsomsin. Energy and emissions impacts of a freeway-based dynamic eco-driving system. 
   Transportation Research Part D: Transport and Environment, {\bf 14}, No. 6,   400--410 (2009).
 	
    
  
  \bibitem{Koller} 
  Koller, M., P. Hemmerle, H. Rehborn, G. Hermanns, B.S. Kerner, and M. Schreckenberg.
            Increased Consumption in Synchronized Flow in Oversaturated City Traffic. In
            Proceedings 10th ITS European Congress, Helsinki, Finland, 2014.
            
  \bibitem{Hemmerle2}
 Hemmerle, P., M. Koller, H. Rehborn, G. Hermanns, B.S. Kerner, M. Schreckenberg.
            Increased Consumption in Oversaturated City Traffic Based on Empirical Vehicle Data. In
            Advanced Microsystems for Automotive Applications 2014. Smart Systems for Safe, Clean
            and Automated Vehicles,
Springer, Cham, Heidelberg, New York, Dordrecht, London,
            2014, pp. 71--79.
            
            
            \bibitem{Hemmerle3}
       Hemmerle, P., H. Rehborn, G. Hermanns, B.S. Kerner, M. Schreckenberg, M. Koller.
            Classification of Empirical Urban Traffic Patterns. In
            Proceedings 10th ITS European Congress, Paper Nr. TP 0008, Helsinki, Finland, 2014.
            
            	 	\bibitem{Kerner_City}
  

 Kerner, B.S.. P. Hemmerle, M. Koller, G. Hermanns, S.L. Klenov, H. Rehborn,   
  and M. Schreckenberg. Empirical synchronized flow in oversaturated city traffic.
 Phys. Rev. E.  {\bf 90}, 032810 (2014).
  
 \bibitem{Kerner}          
            Kerner, B.S.  Verfahren zur Berechnung und Anwendung der
            kumulierten Beschleunigung eines Fahrzeuges pro eine Strecke "j" des Straßennetzes.
           Report of project UR:BAN,   March, 2014 (unpublished).
            
  
\bibitem{Hemmerle}

Hemmerle, P.,  G. Hermanns, M. Koller,   H. Rehborn,  B.S. Kerner, and M. Schreckenberg.
  Estimation of Traffic Dependent Fuel Consumption on Urban Roads.  
  In Proceedings of Transportation Research Board 2015 Annual Meeting, TRB Paper 15-2751  (Washington DC, 2015) (accepted for presentation).
   
    
 
 

  
  
 \bibitem{Hermanns}
Hermanns, G., P.
Hemmerle,   I.N. Kulkov, H. Rehborn,  M. Koller,     B.S. Kerner, and M. Schreckenberg.
  Microscopic Simulation of Synchronized Flow in Oversaturated City Traffic: Effect of Driver's Speed Adaptation.  
  In Proceedings of Transportation Research Board 2015 Annual Meeting, TRB Paper 15-3051   (Washington DC, 2015) (accepted for presentation).
  
  \bibitem{URBAN}
See information about project UR:BAN on  http://urban-online.org/de/urban.html
	 
 	  \bibitem{Rehborn}
Rehborn, H., M. Schreckenberg, B.S. Kerner, G. Hermanns,  P.
Hemmerle,   I.N. Kulkov,  O. Kannenberg, S. Lorkowski, N. Witte, H. B{\"o}hme, T. Finke, and P. Maier. 
Eine methodische Einf{\"u}hrung zur
antriebartabh{\"a}ngigen Routensuche in einem Ballungsraum.
Stra{\ss}enverkehrstechnik, Nr. 3, 151--157 (2014).
    
\end{thebibliography}
\end{document}